\documentclass[showpacs,twocolumn,floatfix,prl]{revtex4}
\usepackage{graphicx}
\usepackage{color}
\usepackage{psfrag}
\usepackage{epsfig}
\usepackage{bm}
\newcommand{\ket}[1]{| #1 \rangle}

\newcommand{\beq}{\begin{eqnarray}}
\newcommand{\eeq}{\end{eqnarray}}

\newcommand{\op}[2]{| #1 \rangle \langle #2 |}

\begin{document}
\title{Entangled microwave photons from quantum dots}
\author{C. Emary, B. Trauzettel, and C. W. J. Beenakker}
\affiliation{Instituut-Lorentz, Universiteit Leiden, P.O.~Box 9506,
2300 RA Leiden, The Netherlands}
\date{May 26th, 2005}
\begin{abstract}
We describe a mechanism for the production of polarisation-entangled
microwaves using intra-band transitions in a pair of quantum dots.
This proposal relies neither on spin-orbit coupling nor on control over
electron-electron interactions.  The quantum correlation of microwave
polarisations is obtained from orbital degrees of freedom in an
external magnetic field.
We calculate
the concurrence of emitted microwave photon pairs, and show that
a maximally entangled Bell pair is obtained
in the limit of weak inter-dot coupling.
\end{abstract}
\maketitle

Entangled photons at optical frequencies are routinely produced by
non-linear optical effects in macroscopic crystals \cite{man95}.
The use of semiconductor nanostructures to produce these states
promises both a greater frequency range and a closer integration with
quantum electronics.

One way to produce entangled photons is to start with entangled
electrons and then transfer this entanglement \cite{cer04}.  However,
one can also start with non-entangled particles, and most work in this area
has focused on the decay of bi-excitonic states in quantum dots
\cite{ben00,sta03,gyw02,per04}.
In
the original proposal \cite{ben00}, a bi-exciton is formed in a
single dot by electrical pumping,
and then subsequently decays by emitting a
pair of photons
via one of two cascades. The polarisations of these two photons are
linked to the cascade by which the bi-exciton decays, and thus, if
these cascades proceed coherently, one produces
polarisation-entangled photons.
All these proposals \cite{cer04,ben00,sta03,gyw02,per04}
involve {\it inter}-band transitions between valence and conduction bands
of a quantum dot, so that the output photons have frequencies in the
{\it visible} range.

\begin{figure}[t]
\begin{center}
\psfrag{a}{\bf{(a)}} \psfrag{b}{\bf{(b)}} \psfrag{c}{\bf{(c)}}
\psfrag{Dl}{L} \psfrag{Dr}{R} \psfrag{1}{1} \psfrag{2}{2}
\psfrag{3}{3} \psfrag{A*}{$A^*,B^*$} \psfrag{Ag}{$A_G,B_G$}
\psfrag{A}{$A$} \psfrag{B}{$B$} \psfrag{L}{$L$}\psfrag{R}{$R$}

\epsfig{file=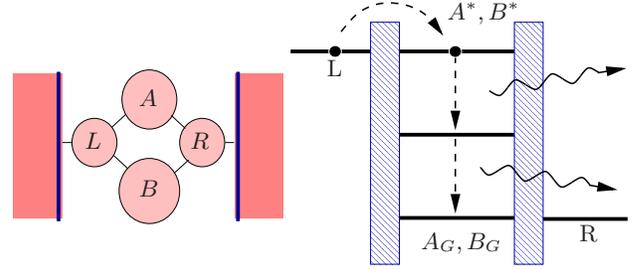, width=0.950\linewidth}
\end{center}
  \caption{
  Schematic of the microwave entangler.
  Left panel: The four-dot arrangement between two electron reservoirs.
  Right panel:  Positions of dot levels,
  with arrows indicating the transitions discussed in the text.
  \label{dots}
  }
\end{figure}

In this paper, we propose the use of {\it intra}-band
transitions of conduction band electrons to generate
entangled {\it microwave} photons.  Our proposal, illustrated in
Fig.~\ref{dots}, can be seen as the real-space analogue of the
bi-exciton decay cascade in energy of Ref. \cite{ben00}.
The microwaves originate from spontaneous
downward transitions between single-particle levels in a quantum dot.
That these transitions couple to microwaves has been demonstrated in
photon-assisted tunnelling experiments \cite{wie02}.

Our entanglement scheme requires four quantum dots
as shown in Fig.~\ref{dots}: two dots ($L$,$R$) to provide unique initial
and final states for the electron, and two more dots ($A$,$B$) to provide the
two decay paths.
For real symmetric tunnel couplings, an electron tunnels through the single level in dot $L$ into
an equal superposition
$2^{-1/2}(\ket{A^*}+ \ket{B^*})$ of upper levels in dots $A$ and $B$.
It decays to the ground state with the emission of two photons. The dots
are configured such that the same process gives rise to two left-circularly-polarized (CP+) photons in dot
$A$ and two right-circularly-polarized (CP$-$) photons in dot $B$. (We will describe later how this can be done.)
The resulting state
$\ket{\Psi}=2^{-1/2}(\ket{A_\mathrm{G}}\ket{++}+\ket{B_\mathrm{G}}\ket{--})$
encodes the state of the quantum dot onto pairs of photons with
left or right circular polarisation.

To disentangle the photons from the electrons we couple dots $A$ and
$B$ symmetrically to a fourth dot $R$, drained by an electron
reservoir. This construction effectively projects the electronic
state in dots $A$ and $B$ onto the even combination
$\ket{\Phi_{+}}=2^{-1/2}(\ket{A_{G}}+\ket{B_{G}})$, since
destructive interference prevents the odd combination
$\ket{\Phi_{-}}=2^{-1/2}(\ket{A_{G}}-\ket{B_{G}})$ from tunneling
coherently into dot $R$. Defining also the even and odd combinations
$\ket{\Psi_{\pm}}=2^{-1/2}(\ket{++}\pm\ket{--})$ of the photon
states, we may write
$\ket{\Psi}=2^{-1/2}(\ket{\Phi_{+}}\ket{\Psi_{+}}+\ket{\Phi_{-}}\ket{\Psi_{-}})$.
The projection of $\ket{\Psi}$ onto $\ket{\Phi_{+}}$ thus produces
the required entangled photon pair $\ket{\Psi_{+}}$.

Whilst not strictly necessary for the entangling mechanism, it is
helpful to assume that the quantum dots are inserted into a cylindrical
microwave resonator able to support both polarisations at the resonant
frequency $\Omega$.
Recent results for double quantum dots and superconducting transmission line
resonators have shown the possibility of extremely large
dot-microwave couplings \cite{bla04,chi04}.
The cavity ensures that microwave, and not phonon, transitions dominate,
and also serves to counteract any slight non-idealities
in the quantum dot emission frequencies that might otherwise render the two decay paths
distinguishable.
But most importantly, the cavity allows operation of the device without postselection.

In the absence of the cavity, it is necessary to detect whether the
electron escapes into the right reservoir after having produced a
photon pair, in order to effectuate the projection onto
$\ket{\Phi_{+}}$. If the electron remains trapped in the state
$\ket{\Phi_{-}}$, the photon pair should be discarded. This
postselection discards one out of two attempts --- even under ideal
conditions. As we will show in what follows, it is possible to
entirely avoid the postselection by using a microwave resonator to
evolve $\ket{\Phi_{-}}\ket{\Psi_{-}}$ into
$\ket{\Phi_{+}}\ket{\Psi_{+}}$. The electron will then always escape
into the reservoir and the required state $\ket{\Psi_{+}}$ is
produced at each and every attempt. The resultant entanglement could
be detected via the violation of a Bell inequality.  This is a
routine experiment for visible light; the analogous experiment at
microwave frequencies is an experimental challenge, with some recent
progress \cite{gab04}.


\begin{figure}[t]
\begin{center}
\psfrag{E1}{$E_1$}
\psfrag{E2}{$E_2$}
\psfrag{E3}{$E_3$}
\psfrag{wa}{$\omega^A_-,$}
\psfrag{wb}{$\omega^B_+,$}
\psfrag{sa}{$\sigma_+$}
\psfrag{sb}{$\sigma_-$}
\psfrag{n}{$n$}
\psfrag{(+)}{CP$+$}
\psfrag{(-)}{CP$-$}
\psfrag{DOT A}{Dot A}
\psfrag{DOT B}{Dot B}
\psfrag{n1}{$1$}
\psfrag{n2}{$2$}
\psfrag{n3}{$3$}
\psfrag{n4}{$4$}
\psfrag{n5}{$5$}
\psfrag{n6}{$6$}
\epsfig{file=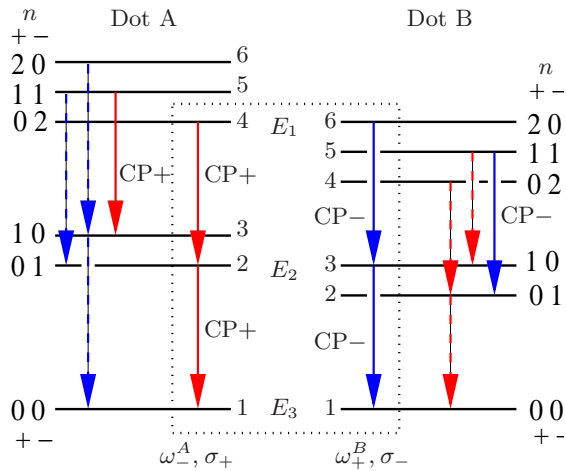, width=0.850\linewidth, height = 6cm}
\end{center}
  \caption{
  Fock-Darwin spectrum of dots $A$ and $B$: the six lowest
  levels with their  quantum numbers $n_\pm$ are shown for each dot.
  The dot sizes are tuned such that
  $\omega_-^A = \omega_+^B=\Omega$.  Selection rule allowed transitions
  are shown, with solid (dashed) arrows denoting on (off) resonant transitions.
  The circular polarisation, CP$+$ or CP$-$,
  of the emitted microwaves is also shown.
  In the entangler,
  electrons are injected into the two levels at $E_1$, and decay via the
  transitions shown in the dotted rectangle.
  In dot $A$, this produces a cascade of two CP$+$
  photons, whereas in dot $B$ two CP$-$  photons are produced.
  \label{lvl}
  }
\end{figure}
After these qualitative considerations, we now turn to a
quantitative description.
The two emitting quantum dots,
$A$ and $B$, are assumed to have cylindrically symmetric,
parabolic confining potentials.
A perpendicular field $B_0$ is applied to control
the microwave polarisations.
The dots are set in a small area such that the charging
energy is sufficient to prevent
more than one electron being in the four-dot system at any time.
Dots $A$ and $B$ are in tunnelling contact with
the single levels in dots $L$ and $R$.
The state $E_L$ of dot $L$ is aligned with the levels $E_1$ in
dots $A$ and $B$,
and the single level $E_R$ in dot $R$ is aligned with
the ground states at $E_3$ (cf.\ Fig.\ \ref{lvl}). The
chemical potentials of the electron reservoirs
are adjusted such that on the left
$\mu_L\gg E_L$ and on the right $\mu_R\ll E_R$.

Since we never occupy the dots
with more than one electron at a time, a spinless single-particle picture
is appropriate. The Hamiltonian of quantum dot $Y=A,B$ is
\beq
  H_Y  = \omega^Y_+ {a^Y_+}^\dag a^Y_+  +\omega^Y_- {a^Y_-}^\dag a^Y_-
 = \sum_i \varepsilon^{Y}_i \op{Y_i}{Y_i}.
   \label{HFD}
\eeq
The excitation energies are $\omega^Y_\pm =\sqrt{{\omega^Y_0}^2
+\omega_c^2/4} \pm \omega_c/2$, with $\omega_c= eB_0/m$ the
cyclotron frequency and $\omega^Y_0$ the confinement energy of dot
$Y$ (we set $\hbar=1$). This so-called Fock-Darwin spectrum
\cite{kow01} is thus determined by the two quantum numbers $n^Y_\pm$
\cite{FDnote}. Electric dipole transitions between dot levels with
photon emission along the symmetry axis satisfy the selection rule
$|\Delta n_\pm|=1$. Transitions in which $n_-$ decreases emit CP$+$
photons, and transitions in which $n_+$ decreases emit CP$-$
photons.


Figure \ref{lvl} shows the six lowest levels $\ket{Y_i}$ in dots $Y=A,B$,
labelled $i=1$ to 6 from the bottom up.  We take $B_0$
small enough that there are no
level crossings in the spectrum.
Since $\omega_\pm$ depends on the confinement energy $\omega_0$,
the spectra of
the dots can be tuned by electrostatically changing their sizes.
We set $\omega_-$ of dot $A$ equal to $\omega_+$ of dot $B$ by choosing
the ratio of the linear sizes to be
$l_A/l_B \approx 1- l_B^2/2l_c^2$ for $l_c\gg l_B$,
with $l_c = \sqrt{\hbar/eB_0}$ the magnetic length.
Fixing $\omega_-^A = \omega_+^B=\Omega$, only one free parameter
$\omega_c \ll \Omega$ remains to specify the two spectra.


We calculate the coupled dynamics of electron and photons
from the master equation \cite{sto96,gur98,bra04},
\beq
  \frac{d}{dt}\rho  =-i \left[H,\rho \right]
  + {\cal L} \left[\rho\right].
 \label{ME}
\eeq
The commutator describes the coherent evolution of the density
matrix $\rho$ under the action of the Hamiltonian $H$ of the
dot-cavity system.  The operator ${\cal L} \left[\rho\right]$
describes the coupling of dot $R$ to the electron reservoir on the
right. As we are interested in the entanglement produced by the
passage of a single electron through the device, we initialise the
system with an electron in dot $L$.  The left reservoir serves only
to populate this level initially and is then decoupled, while the
coupling to the right reservoir is permanent and acts as a sink for
the electron.

The Hamiltonian is
$H = \sum_Y H_Y + H_T + H_+ + H_- + H_{\mu}$, with the sum $Y$ taken
over all four dots.  The Hamiltonians of the dots $A$ and $B$
contain six levels each, according to Eq. (\ref{HFD}).
For dots $L$ and $R$ we have $H_L=E_L\op{L}{L}$ and $H_R=E_R\op{R}{R}$.
We set $E_L=E_1$ and $E_R=E_3$.
The dots are connected via the tunnelling Hamiltonian
\beq
  H_T &=& \sum_{i=1}^6
  \bigl\{
    T_{LA_i} \op{L}{A_i} + T_{RA_i} \op{R}{A_i}
  ~~~~ ~~ ~~~~~~ ~~~~ ~~
  \nonumber
  \\
  &&~~~~
    + T_{LB_i} \op{L}{B_i} + T_{RB_i} \op{R}{B_i}
  \bigr\}
  + \mathrm{H.c.},
\eeq
where $T_{XY_i}$ are tunnel amplitudes.

The microwave photons have the Hamiltonians
$H_+ = \omega^A_- b_+^\dag b_+$ and $H_- = \omega^B_+ b_-^\dag b_-$,
with $b_\pm$ the field operators of CP$\pm$ microwaves.
Since we are on resonance, $\omega_-^A=\omega_+^B=\Omega$.
In the rotating wave approximation, the emission of microwaves is
governed by the Hamiltonian
\beq
H_{\mu}
 &=&
  g^A
  \bigl\{
    \op{A_3}{A_5} + \op{A_2}{A_4}  + \op{A_1}{A_2}
   \bigr\} b^\dag_{+}
  ~~~~~~~~~~~~~
 \nonumber
  \\
  &&
\!\!\!\!\!\!\!\!\!\!
  \mbox{}
  +
  g^B
  \bigl\{
    \op{B_3}{B_6} + \op{B_2}{B_5}  + \op{B_1}{B_3}
   \bigr\} b^\dag_{-}
  +\mathrm{H.c.},
\eeq
neglecting off-resonant transitions.

\begin{figure}[t]
\begin{center}
\psfrag{t}{$ \Omega t$}
\psfrag{C}{${\cal C}$}
\psfrag{N2}{$N_2$}
\epsfig{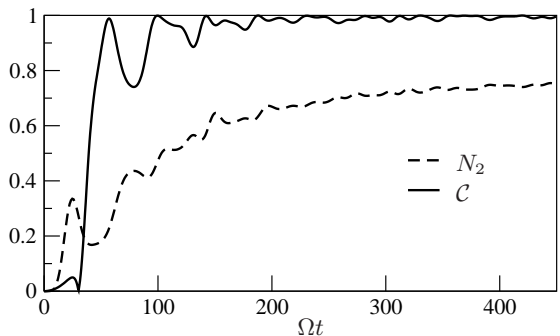}
\end{center}
  \caption{
  Dynamics of the microwave entangler.
  Plotted is the time-dependence of the mean number of photon
  pairs in the cavity $N_2$ and their degree of entanglement
  ${\cal C}$ (the concurrence).
  For long times, the field approaches a steady state with,
  for the weak coupling $T_A=T_B=0.05\,\Omega$ shown here,
  a high proportion of photon pairs $N_2^\infty \approx 0.7$
  and almost maximal concurrence ${\cal C}^\infty \approx 1$.
  Model parameters were $\omega_c=g=\Gamma_R=0.1\,\Omega$.
  \label{res}
  }
\end{figure}

\begin{figure}[t]
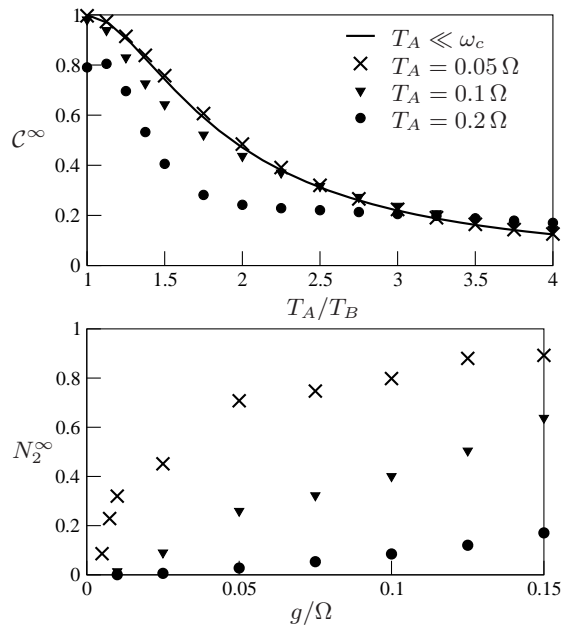

\begin{center}
\psfrag{Cinf}{${\cal C}^\infty$} \psfrag{TT}{$T_A/T_B$}
\psfrag{TA=0.05}{$T_A=0.05\,\Omega$}
\psfrag{TA=0.1}{$T_A=0.1\,\Omega$}
\psfrag{TA=0.2}{$T_A=0.2\,\Omega$} \psfrag{ideal}{$T_A\ll \omega_c$}
\psfrag{Ninf}{$N_2^\infty$} \psfrag{g}{$g/\Omega$}
\psfrag{T=0.05}{$T=0.05$} \psfrag{T=0.1}{$T=0.1$}
\psfrag{T=0.2}{$T=0.2$} \epsfig{file=./r_c.eps,
clip=true,width=0.85\linewidth} \epsfig{file=./r_N2.eps,clip=true,
width=0.85\linewidth}
\end{center}
  \caption{
  Performance of the microwave entangler.
  Upper panel:  The asymptotic entanglement of the photon pairs
  as a function of the asymmetry in the couplings
  of dots $A$ and $B$ to the other two dots.
  The solid curve is the analytic result (\ref{ideal}) for
  $T_A,T_B \ll \omega_c$.  We fixed $g=\omega_c$.
  Lower panel: Probability $N_2^\infty$ of successful
  operation of the entangler
  as a function of the dot-microwave coupling $g$.  The symbols
  correspond to the same values of $T_A$ as in the upper panel and
  we fixed $T_A=T_B$.
  In both plots we took $\omega_c=\Gamma_R=0.1\,\Omega$.
  \label{res2}
  }
\end{figure}

The coupling of the right electron reservoir
to dot $R$ is incorporated into the
master equation through the Lindblad operator
$
{\cal L}\left[\rho\right]
  =  D \rho D^\dag - \frac{1}{2} D^\dag D \rho
  - \frac{1}{2} \rho D^\dag D
$.
The jump operator is $D= \sqrt{\Gamma_R}\op{0_\mathrm{dot}}{R}$,
with $\Gamma_{R}$ the tunnelling rate  and
$\ket{0_\mathrm{dot}}$ denoting all four dots empty.
In the following, we will for simplicity set $g^A=g^B=g$ and
$T_{XA_i}=T_A$, $T_{XB_i}=T_B$ for $X=L,R$ and $i=1, \cdots ,6$.
Results of a numerical integration of the master equation (\ref{ME})
are plotted in Figs.~\ref{res} and \ref{res2}.


The density matrix of the field is obtained by tracing
out the dot degrees of freedom from the density matrix $\rho(t)$.
We denote its (unnormalised) two-photon projection as $\chi(t)$.  It has
the form
\begin{eqnarray}
   \chi &=&
   r_{+}\op{++}{++}
   + r_{-}\op{--}{--}
  \nonumber\\
   &&
   \mbox{}
  + r_{c}\op{++}{--} +  r_{c}^*\op{--}{++}.
\end{eqnarray}
The mean number of photon pairs in the cavity is given by
$N_2 = \mathrm{Tr}~\chi=r_++r_-$.  This is plotted as a function of time
in Fig. \ref{res}.
The time scale on which
the electron is transmitted sequentially
through the elements of the system is
$\tau = \Gamma_R^{-1} + 2 g^{-1} +  T_A^{-1} + T_B^{-1}$.
We see that for times $t \gg \tau \approx 70\,\Omega^{-1}$,
the number of photon pairs $N_2$
approaches a stationary value, $N_2^\infty$.

The degree of entanglement (concurrence) ${\cal C}$ of the photon pair
can be calculated using  Wootter's formula \cite{woo98} for the
concurrence of the density matrix $\chi(t)$, which in general
describes a mixed state.  We find
\beq
  {\cal C}=\frac{2|r_{c}|}{r_++r_-}
  \label{conc}.
\eeq
The time-dependent concurrence ${\cal C}(t)$ is shown in
Fig.~\ref{res}, and this is seen to saturate for $t \gg \tau$. In
Fig. \ref{res2} (upper panel) we plot this long-time limit ${\cal
C}^\infty$ as a function of the coupling asymmetry $T_A/T_B$ for
several values of $T_A$.

An analytical solution is possible for $T_A, T_B \ll \omega_c$.
Since the top three levels in dots $A$ and $B$
are each separated  by $\omega_c$, when $T_A, T_B \ll \omega_c$
the electron tunnels from dot $L$ only into the resonant levels at $E_1$,
producing a photon-pair before leaving the dots.
In this limit, the concurrence ${\cal C}^\infty$ of the final state
is easily calculated, as the relative amplitudes for the generation of each of
the two photon pairs are proportional to the product of the individual
coupling amplitudes along each of the two paths.  We find
\beq
  {\cal C} = 2 \frac{|T_A T_B g_A g_B|^2}{|T_A g_A|^4 + |T_B g_B|^4}
  \label{ideal}.
\eeq
The numerical weak coupling results in Fig. \ref{res2} (crosses),
are very close to this analytic result (solid curve). For weak,
symmetric inter-dot couplings, the concurrence approaches unity,
corresponding to the production of the Bell state $2^{-1/2}
(\ket{++}+\ket{--})$.  The negative effects of level broadening
induced by the inter-dot couplings are slight provided $T_A,T_B
\lesssim \omega_c$.

Successful
operation of the device requires that, after the passage of the electron,
the cavity is left with a pair of microwave photons.  The
probability of successful operation is thus given by the asymptotic
mean number of photon pairs $N_2^\infty$.  This probability is plotted
in Fig. \ref{res2} (lower panel). The success probability tends to unity in the weak-coupling limit.

In the final part of this paper we consider two different mechanisms
by which the efficiency of the device is reduced.  The first
mechanism is direct inelastic transitions from upper levels in dots
$A$ and $B$ into dot $R$.  This reduces the number of photon pairs
produced, but provided that the rate of these inelastic processes is
smaller than the dot-cavity coupling, photon-pair production will
still dominate.  Moreover, provided that these inelastic processes
give no which-way information on the path of the electron, affecting
the two dots roughly equally, they have little effect on the degree
of entanglement of the photon pairs that are emitted.

\begin{figure}[t]
\begin{center}
\psfrag{Cinf}{${\cal C}^\infty$} \psfrag{TT}{$T_A/T_B$}
\psfrag{lg=0.05}{$\Gamma_{\phi}=0.05\,\Omega$}
\psfrag{lg=0.01}{$\Gamma_{\phi}=0.01\,\Omega$} \psfrag{g}{$g/\Omega$}
\epsfig{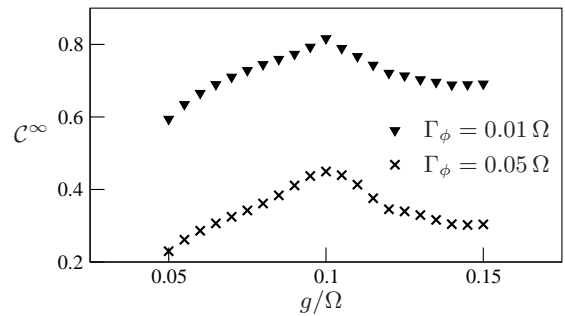}
\end{center}
  \caption{ Effect of decoherence, with rate $\Gamma_{\phi}$, on the asymptotic entanglement of the photon pairs.
  We fixed $T_A=T_B=0.05\,\Omega $, $\omega_c=\Gamma_R=0.1\,\Omega$ and varied $g$.
  \label{res3}
  }
\end{figure}

The second mechanism is decoherence of the two spatially separated
decay paths. The decoherence typically results when the charge on
one of the dots couples to other charges in the environment, thereby
providing which-way information. We model this charge noise by
adding to the master equation (\ref{ME}) jump operators
\begin{equation}
D_{Y}=\sqrt{\Gamma_{\phi}} \left(\sum_{i}|Y_{i}\rangle\langle
Y_{i}|\right)\otimes\openone_{\rm photon}
\end{equation}
which measure the charge on each of the four dots $Y=A,B,L,R$.  Here
the sum over $i$ ranges over all states $|Y_{i}\rangle$ in quantum
dot $Y$, the symbol $\openone_{\rm photon}$ denotes the identity for
the photon degrees of freedom, and $\Gamma_{\phi}$ is the
decoherence rate. The resulting degradation of the concurrence is
plotted in Fig.\ \ref{res3}, for two values of $\Gamma_{\phi}$. More
extensive data indicates that the degradation is algebraic, ${\cal
C}\propto (\tau^{\ast}\Gamma_{\phi})^{-1}$, with $\tau^{\ast}$ a
characteristic time scale of the device that takes on it smallest
value when all transition rates $g,T_{A},T_{B},\Gamma_{R}$ are close
to each other. In Fig.\ \ref{res3} we vary the electron-photon
coupling constant $g$, and indeed observe that the concurrence is
maximized when $g$ is neither much smaller nor much larger than the
other rates. Because the degradation of the concurrence is only
algebraic, rather than exponential, a reasonable amount of charge
noise can be tolerated.

In summary, we have described a mechanism for the production of
entangled microwave photon pairs using intra-band transitions in
quantum dots.   Our calculations indicate that a four-dot device is
capable of the output of highly entangled pairs with useful success
probability.  The entangler may be thought of as an electron
interfer\-ometer in which each of the two paths is coupled to a
different photon-pair producing process. Apart from the different
frequency range, it differs from the celebrated bi-exciton entangler
\cite{ben00} in that the interfering paths are in real space, rather
than in energy space, and also in that the correlated polarisations
result from orbital selection rules --- rather than from spin-orbit
coupling.

This work was supported by the Dutch Science Foundation NWO/FOM.



\begin{thebibliography}{99}
  \bibitem{man95} L. Mandel and  E. Wolf,
  {\it Optical Coherence and Quantum Optics}  (Cambridge, 1995).
  \bibitem{cer04}  V. Cerletti, O. Gywat, and D. Loss, cond-mat/0411235;
  M. Titov, B. Trauzettel, B. Michaelis, and C.W.J. Beenakker,
  cond-mat/0503676.
  \bibitem{ben00} O. Benson, C. Santori, M. Pelton, and Y. Yamamoto,
  Phys. Rev. Lett. {\bf 84}, 2513 (2000).
  \bibitem{gyw02}  O. Gywat, G. Burkard, and D. Loss,
    Phys. Rev. B {\bf 65}, 205329 (2002).
  \bibitem{sta03} T. M. Stace, G. J. Milburn, and C. H. W. Barnes,
    Phys. Rev. B {\bf 67}, 085317 (2003).
  \bibitem{per04} J. I. Perea and C. Tejedor,  cond-mat/0409745.
  \bibitem{wie02}  W.G. van der Wiel, S. de Franceschi, J. M. Elzerman,
    T. Fujisawa, S. Tarucha, and L. P. Kouwenhoven,
    Rev. Mod. Phys. {\bf 75}, 1 (2003).
  \bibitem{bla04} A. Blais, R.-S. Huang, A. Wallraff,
    S. M. Girvin, and R. J. Schoelkopf,  Phys. Rev. A {\bf 69}, 062320 (2004).
  \bibitem{chi04}  L. Childress, A. S. S\o rensen, and M. D. Lukin,
  Phys. Rev. A {\bf 69}, 042302 (2004).
  \bibitem{gab04} J. Gabelli, L.-H. Reydellet, G. F\`eve,
    J.-M. Berroir, B. Pla\c cais, P. Roche, and D. C. Glattli,
    Phys. Rev. Lett. {\bf 93}, 056801 (2004).
  \bibitem{kow01}  L.~P.~Kouwenhoven, D.~G.~Austing, and S.~Tarucha,
    Rep.~Prog.~Phys.~{\bf 64}, 701 (2001).
  \bibitem{FDnote} The quantum numbers $n^Y_\pm$ are related to the set
    $\left\{n,l\right\}$ of Ref. \cite{kow01} via $n = \mathrm{Min}(n_+,n_-)$
    and $l = n_- - n_+$.
  \bibitem{sto96} T. H. Stoof and Yu. V. Nazarov, Phys. Rev. B {\bf 53},
    1050 (1996).
  \bibitem{gur98} S. A. Gurvitz, Phys. Rev. B {\bf 57}, 6602 (1998).
  \bibitem{bra04} T. Brandes, Phys. Rep. {\bf 408}, 315 (2005).
  \bibitem{woo98} W. K. Wootters, Phys. Rev. Lett. {\bf 80}, 2245 (1998).
\end{thebibliography}
\end{document}